\documentclass{article}
\usepackage{spconf,amsmath,amssymb,graphicx,xcolor,url,cite}
\usepackage{booktabs,comment}
\usepackage{here}

\title{Attention-based Multi-hypothesis Fusion for Speech Summarization}
\name{Takatomo Kano$^1$, Atsunori Ogawa$^1$, Marc Delcroix$^1$, and Shinji Watanabe$^2$}
\address{$^1$NTT Corporation, Japan\\
$^2$Language Technologies Institute, Carnegie Mellon University, Pittsburgh, USA}
\begin{document}
\ninept
\maketitle
\vspace{-5pt}
\begin{abstract}
\vspace{-5pt}
Speech summarization, which generates a text summary from speech, can be achieved by combining automatic speech recognition (ASR) and text summarization (TS). With this cascade approach, we can exploit state-of-the-art models and large training datasets for both subtasks, i.e., Transformer for ASR and Bidirectional Encoder Representations from Transformers (BERT) for TS. However, ASR errors directly affect the quality of the output summary in the cascade approach. We propose a cascade speech summarization model that is robust to ASR errors and that exploits multiple hypotheses generated by ASR to attenuate the effect of ASR errors on the summary. We investigate several schemes to combine ASR hypotheses. First, we propose using the sum of sub-word embedding vectors weighted by their posterior values provided by an ASR system as an input to a BERT-based TS system. Then, we introduce a more general scheme that uses an attention-based fusion module added to a pre-trained BERT module to align and combine several ASR hypotheses. Finally, we perform speech summarization experiments on the How2 dataset and a newly assembled TED-based dataset that we will release with this paper\footnote{https://github.com/nttcslab-sp-admin/TEDSummary}. These experiments show that retraining the BERT-based TS system with these schemes can improve summarization performance and that the attention-based fusion module is particularly effective.
\vspace{-5pt}
\end{abstract}
\vspace{-10pt}
\section{Introduction}
\label{sec:intro}
\vspace{-8pt}
Speech summarization generates a text summary from given speech data. It is challenging because it needs to process lengthy speech data (a sequence of utterances) and extract important information to create a compact representation of the content. Moreover, in contrast to a text input, speech contains fillers, disfluencies, redundancies (e.g. repetition of the same phrases), and colloquial language. There are two main types of summarization approaches, \emph{extractive} and \emph{abstractive}. Extractive summarization aims at identifying the most relevant segments of the input text/speech document and then concatenating them to assemble a summary. Abstractive summarization aims at directly generating a summary by paraphrasing the intent of the input document. Although the latter type is challenging, it has achieved great progress with the introduction of powerful deep learning models for text summarization (TS) such as Bidirectional Encoder Representations from Transformers (BERT)~\cite{DBLP:conf/naacl/DevlinCLT1}. Moreover, abstractive summarization can potentially normalize spoken text to remove disfluencies, redundancies, and colloquial language, making the summary more understandable than extractive ones. Consequently, in this paper, we focus on abstractive summarization.

Speech summarization is achieved by combining two main sub-modules: an automatic speech recognition (ASR) module, which transcribes speech into a corresponding text document, and a TS system, which generates a compact representation of that document. Such a cascade connection permits using state-of-the-art modules optimized for each task individually, without requiring a large amount of paired data composed of speech data and associated summaries. Moreover, each module can operate on very different time resolution, i.e., ASR is performed for each utterance, while TS requires the entire text document. However, while generating the summary, a cascade connection discards speech-specific information~\cite{DBLP:conf/interspeech/KanoTSNTN13,DBLP:journals/taslp/DoSN18}, which has the potential to enrich the summary, such as intonation when generating the summary. Moreover, the TS system receives input text containing ASR errors that affect the performance of the speech summarization system~\cite{DBLP:conf/eusipco/WengLC20}. This study focuses on the latter problem.

Many works have attempted to mitigate the influence of ASR errors on a natural language processing (NLP) back-end.  For example, studies on speech translation~\cite{DBLP:conf/icassp/BertoldiZF07,DBLP:conf/emnlp/HopkinsM11,DBLP:conf/interspeech/OhgushiNSTN13,DBLP:journals/corr/kaho,DBLP:conf/slt/BaharBSN21,DBLP:conf/emnlp/SperberNNW17} have reported that ASR errors could be mitigated during translation by considering multiple recognition hypotheses at the input of the translation back-end. Some studies~\cite{DBLP:journals/corr/kaho,DBLP:conf/slt/BaharBSN21} have used posterior probabilities to weight ASR hypotheses based on their confidence. Sperber et al.~\cite{DBLP:conf/emnlp/SperberNNW17} proposed directory inputting recognition lattices to the back-end translation system. In general, the recognition lattices hold the approximated entire ASR search information representing word-level multi-hypotheses with a compact lattice form.
Therefore, this approach helps mitigating ASR errors by considering alternative word candidates within the recognition lattices. However, it is difficult to use this approach directly with pre-trained state-of-the-art TS models like BERT, since the BERT model expects a sub-word sequence as an input instead of a lattice. For speech summarization, some studies built systems that are robust to ASR errors~\cite{DBLP:conf/icassp/OgawaHNN19,DBLP:journals/taslp/XieL11,DBLP:conf/interspeech/LinC09}. Weng et al.~\cite{DBLP:conf/eusipco/WengLC20} achieved robust speech summarization by adding confidence scores associated with each recognized word to the input of a BERT-based TS system. The confidence scores help the TS system to ignore unreliable words in the ASR output. However, it provides a limited ability to recover the unreliable information from alternative word candidates as it derives from a single (the 1-best) ASR hypothesis rather than an N-best list or a lattice. Ogawa et al.~\cite{DBLP:conf/icassp/OgawaHNN19} proposed inputting confusion networks (CNs) to a compressive (i.e., non-neural) TS system. Then, from the CNs, the TS system selects recognized words to form a summary that maximizes the ILP-based objective function. These studies confirmed that inputting multiple ASR hypotheses (e.g. lattices and CNs) and auxiliary information (e.g. confidence scores) to the NLP module is useful~\cite{DBLP:conf/emnlp/SperberNNW17,DBLP:journals/corr/kaho,DBLP:conf/interspeech/OhgushiNSTN13}.
 
In this paper, we propose a speech summarization model that can exploit multiple ASR hypotheses to mitigate the influence of ASR errors and be used with pre-trained TS systems like BERT. We explore two approaches to combining the ASR hypotheses. First, we propose replacing the input sub-word embedding of BERT with a sum of sub-word embedding vectors weighted by their ASR posterior values, as done in previous studies on speech translation~\cite{DBLP:journals/corr/kaho}. The second approach proposes to use an attention fusion mechanism to combine different ASR hypotheses within the BERT module. In the point of fusing multiple inputs with an attention mechanism, this attention fusion is similar to hierarchical attention~\cite{DBLP:conf/naacl/YangYDHSH16,DBLP:conf/acl/LibovickyH17} proposed for multi-stream combinations~\cite{DBLP:conf/icassp/WangLMHWH19} and audio-visual processing~\cite{DBLP:conf/iccv/HoriHLZHHMS17}. However, our proposal does not have a hierarchical architecture; it is an attention function that fuses multiple-hypothesis following the self-attention manner. The query is a 1-best hypothesis, and value and key are multiple-hypothesis. The attention fusion can be implemented at the input or within the BERT model. The latter can exploit BERT's strong modeling capability to perform the hypotheses fusion. 

Posterior fusion is similar to the confidence-based approach~\cite{DBLP:conf/eusipco/WengLC20} in the sense that both approaches simply exploit the confidence of individual words or tokens within the 1-best hypothesis at token level and do not explicitly model multiple hypotheses at sequence level. In contrast, the attention fusion explicitly models multiple token sequence hypotheses by using two attention steps. First, we align the hypotheses with the 1-best hypothesis using an attention mechanism across the tokens of the 1-best and each hypothesis. This process allows using hypotheses with different lengths or redundant information. In the second step, we employ an attention mechanism over the aligned tokens and across the hypotheses to combine them. This attention fusion is similar to system combination approaches like ROVER~\cite{ROVER}, posterior probability decoding~\cite{Evermann} and minimum Bayes risk decoding~\cite{GOEL2000115}, but the combination is performed within the BERT encoding process.

We performed experiments to confirm the effectiveness of the proposed methods on two speech summarization datasets, i.e., YouTube How2 video and TED Talk summarization. The TED Talk summarization is a newly assembled corpus that associates the TEDLIUM ASR corpus with publicly available TED Talk summaries. We release this new corpus with this paper. Experimental results show that retraining a BERT-based TS system with the proposed multi-hypothesis combination schemes can improve summarization performance on both datasets.

\vspace{-10pt}
\section{Speech summarization}
\vspace{-5pt}
Let us consider a spoken document $D$, which contains $K$ utterances. Let $X_k$ and $S_k$ be the speech signal and the associated transcription of the $k$-th utterance of this spoken document. The summarization task consists of generating a compact document $Y$ from the input spoken document $D$.
This problem is addressed in two stages. First, we use an ASR system to transcribe each speech utterance $X_k$ into text. Then, we use a TS to generate the summary $Y$ from the input speech document $D$.
\vspace{-10pt}
\subsection{ASR system}
\label{sec:ASR}
\vspace{-5pt}
We use a state-of-the-art Transformer model for ASR~\cite{DBLP:conf/asru/KaritaWWYZCHHIJ19,DBLP:conf/acl/InagumaKDKYHW20}. The ASR system predicts the word sequence, $\hat{S}_k$ associated with the speech signal $X_k$ using beam search. This is achieved by combining the scores from the transformer and language model,
\begin{align}
\hat{S}_k &= \mathrm{argmax}_{S}\left(\log p_{\text{transf}}(S|X_k)+ \lambda \log p_{\text{lm}}(S)\right),
\end{align}
where $p_{\text{transf}}(S|X_k)$ is the posterior probability of $S$ given $X_k$ obtained with the transformer model,  $p_{\text{lm}}(S)$ represents the language model, and $\lambda$ is the shallow fusion weight of the language model.

In this paper, we assume that we can obtain multiple ASR hypotheses. These hypotheses can be the $N$-best recognition hypotheses obtained using beam search decoding or 
$N$ hypotheses obtained with different ASR systems as often done with system combination~\cite{DBLP:journals/csl/JalalvandNFMT18}. In the following, $\hat{S}^n_k$ represents the $n$-th hypothesis obtained for a list of $N$ hypotheses.

In practice, ASR operates on sub-word units such as byte pair encoding (BPE), and thus a recognition hypothesis can be expressed as $\hat{S}^n_k = [\hat{s}^n_{k,1},\ldots,\hat{s}^n_{k,M_k^n}]^{\mathrm{T}} \in \mathbb{R}^{M_k^n\times V}$, where $\hat{s}^n_{k,m}$ is a one-hot vector representing the  $m$-th token of the $n$-th hypothesis and the $k$-th utterance,  $M_k^n$ is the length of the hypothesis, $V$ is the vocabulary size (the number of sub-word units), and $\mathrm{T}$ is the transpose operation.
To obtain recognition hypotheses for the entire spoken document, we can simply concatenate the hypotheses of all utterances as $\hat{S}^{n}=[\hat{s}^n_{1,1},\ldots,\hat{s}^n_{K,M_K^n}]^{\mathrm{T}}$. By abuse of notation, we remove the utterance index $k$ and redefine $\hat{S}^n=[\hat{s}^n_{1},\ldots,\hat{s}^n_{M^n}]^{\mathrm{T}}\in \mathbb{R}^{M^n\times V}$, where $M^n$ is the total length of the concatenated $n$-th hypotheses of the the spoken document, i.e, $M^n = \sum_{k} M^n_k$. With beam search, we can also obtain the posterior probabilities associated with each token in the hypothesis, which we denote as $\hat{p}^n_{m}$.
\vspace{-5pt}
\subsection{TS system}
\vspace{-5pt}
Early works on text summarization used combinatorial optimization approaches to find important content in a text document~\cite{DBLP:conf/ecir/McDonald07,DBLP:conf/ijcai/YihGVS07,DBLP:conf/cikm/TakamuraO09}. It was difficult to generate consistent abstractive summaries with such approaches,  so most research focused on extractive summaries.
The success of the deep learning-based language models greatly improved the quality of abstractive summarization~\cite{DBLP:conf/icml/ZhangZSL20,DBLP:conf/acl/SeeLM17,DBLP:conf/icml/ZhangZSL20,DBLP:conf/emnlp/LiuL19}. Recently, Transformer-based models have become state-of-the-art models for abstractive text summarization~\cite{9328413}. For example, BERTSum~\cite{DBLP:conf/emnlp/LiuL19} leverages the strong language modeling capability of a pre-trained BERT \cite{DBLP:conf/naacl/DevlinCLT1} model to achieve high-quality abstractive summarization through transfer learning. Since, the original BERT model assumes single sentences as input instead of a sequence of sentences as in summarization tasks, BERTSum introduces BERT's classification (CLS) tokens at the start of each sentence of the input document. BERTSum uses the BERT model as an encoder and adds a Transformer-based decoder to generate the summary. The model is then fine-tuned on the text summarization task.

The TS system accepts the entire transcription of the spoken document as an input. In general, we can simply use the 1-best hypothesis $\hat{S}^1$. We employ a BERTSum model to generate a summary from $\hat{S}^{1}$ as
\begin{align}
E &= \text{Emb}(\hat{S}^{1}), \label{eq:embbedding} \\
Z &= \text{Enc}^{\text{bert}}(E), \\
Y &= \text{Dec}^{\text{transf}}(Z),
\end{align}
where $\text{Emb}(\cdot)$ is an embedding layer that converts the one-hot token sequence into a sequence of embedding vectors $E=[e_1,\ldots,e_{M^1}]^{\mathrm{T}} \in \mathbb{R}^{M^1\times B}$, $B$ is the size of the embedding, $\text{Enc}^{\text{bert}}(\cdot)$ represents a BERT encoder, $Z$ is an intermediate representation of the document, and $\text{Dec}^{\text{transf}}(\cdot)$ represents a transformer-based decoder that generates a summary based on $Z$.
BERTSum takes advantage of the strong language modeling capability of the pre-trained BERT model to generate high-quality summaries. By inserting CLS symbols between consecutive sequences, the BERT encoder can process a sequence of multiple utterances that forms a long document.

There are two issues with the cascade connection of ASR and BERTSum. First, the BERT model assumes discrete features representing the IDs of the sub-word units as inputs. Thus it cannot accept an uncertain input such as posteriors of the ASR system directly, and recognition errors will directly affect the summary. We discuss some proposals to address this problem in  Section~\ref{sec:proposal}.

Second, although both systems use sub-word units, the definition of sub-word units usually differs. Typically, ASR performs better with much smaller BPE units than the conventional BERT system.. However, since we would like to exploit a pre-trained BERT model, we consider two options. With the first option, we re-tokenize the word sequence after ASR to match the sub-word definition of the BERT model. With this approach, recognition performance may be optimal, but it offers limited possibilities for the interconnection of the ASR and TS systems.
The other option is to use an ASR system trained with the same sub-word unit definitions as the BERT model. This training will degrade ASR performance but allow more flexibility in combining the two systems. We will analyze the impact of these options in the experiments of Section~\ref{sec:bpe size}.
\vspace{-15pt}
\subsection{Conventional interconnection of ASR and TS systems}
\label{sec:confidence}
\vspace{-5pt}
Many ASR+NLP systems such as speech summarization, translation, and spoken dialog systems use a cascade of ASR and NLP sub-systems built independently. The performance of the downstream NLP task is directly affected by the recognition errors~\cite{DBLP:conf/iwslt/2010}. Previous studies improved the robustness of the NLP back-end to ASR errors using various auxiliary information sources from the ASR system, e.g., probabilities, recognition hypotheses, and hidden states~\cite{DBLP:conf/icassp/HeDA11,DBLP:conf/iwslt/KanoSTNTN12,DBLP:conf/interspeech/ManakulGW20,DBLP:conf/interspeech/RuizGLF15,DBLP:journals/corr/kaho,DBLP:conf/icassp/OgawaHNN19,DBLP:conf/slt/KanoS021,DBLP:conf/emnlp/HopkinsM11,DBLP:conf/emnlp/SperberNNW17,DBLP:conf/eusipco/WengLC20,DBLP:conf/naacl/DalmiaYRMW21}.

For speech summarization, Weng et al. proposed including a confidence embedding in the input of BERTSum~\cite{DBLP:conf/emnlp/LiuL19} to achieve robust speech summarization~\cite{DBLP:conf/eusipco/WengLC20}. They modified the input embedding vectors of BERTSum to be the sum of the sub-word embedding vector and a confidence embedding. Here, we implemented a similar method. We use $\hat{p}^1_m$, as introduced in Section~\ref{sec:ASR}, as a confidence value, which we map to a hidden vector using a linear mapping as
\begin{equation}
    c_m = \text{Emb}^{\text{conf}}(\hat{p}^1_m),
\end{equation}
where $c_m \in \mathbb{R}^B$ is a confidence embedding and $\text{Emb}^{\text{conf}}(\cdot)$ is a linear embedding layer, where $B$ is the dimension of the projected embedding vectors. Then the modified embedding vector $e_m^{\text{conf}} $ is obtained as the summation of confidence and word embeddings:
\begin{equation}
 \label{eq:confidence}
     e_m^{\text{conf}} = c_m+e_m.
\end{equation}
\vspace{-10pt}
\section{Multi-hypothesis Summarization}
\vspace{-5pt}
\label{sec:proposal}
In this paper, we propose a BERT-based summarization model that takes into account multiple speech recognition hypotheses. 
We propose combining several ASR hypotheses to mitigate the influence of ASR errors. Conceptually, the combined hypothesis can be obtained as a weighted-sum as,  
\begin{align}
\label{eq:general}
    e_m^* = \sum_{n=1}^N \alpha^n_m e_m^n,
\end{align}
where  $\alpha^n_m$ denotes a weight of each hypothesis, $e_m^*$ denotes a modified embedding vector.

First, we apply a method of incorporating posterior probability that was originally proposed for speech translation to the BERT summarization model. Next, we explain the hypothesis fusion method using an attention mechanism.
\vspace{-5pt}
\subsection{Speech summarization with posterior fusion}
\label{sec:posterior fusion}
\vspace{-5pt}
The posterior fusion consists of summing the embedding vectors of all sub-words weighted by their posterior probabilities, $\hat{p}_m^n$, as
\begin{align}
 \label{eq:posterior}
     e_m^{\text{post}} =& \sum_{n=1}^N \hat{p}_m^n e_m^n,
\end{align}
where $e_m^{\text{post}}$ is a modified embedding vector and $e_m^n$ is the $m$-th embedding vector of the $n$-th hypothesis obtained with Eq. (\ref{eq:embbedding}). 
The modified embedding $e_m^{\text{post}}$ can include the uncertainty in the ASR system. Computing $e_m^{\text{post}}$ requires that all hypotheses are aligned and have the same length. 
We thus create the $N$ hypotheses as follows. First, we save the sequence of sum of output log-softmax values from the ASR and the language models for the best beam-search path. After decoding, for each step in the 1-best path, we select the $N{=}10$ tokens with the top 10 values of saved output vectors in the path. This way of creating $N$ hypotheses may generate more diverse hypotheses than simply obtaining the $N$-best list from beam search decoding. Note that since the posterior-based fusion modifies the input of the TS model, we need to retrain the BERTSum model using ASR hypotheses.

\vspace{-8pt}
\subsection{Attention-based multi-hypothesis fusion}
\label{sec:attention fusion}
\vspace{-5pt}
Posterior probability fusion trusts the ASR weighting and may mitigate the influence of unreliable tokens when performing summarization. 
However, in the case of the ASR outputs the correct word at $10$-th hypotheses, it may be difficult to recover information from the modified embedding of Eq.~(\ref{eq:posterior})  because the correct word's weight $\hat{p}^{10}_{m^10}$ is too small. On the other hand, our proposal re-calculates for all hypothesis weight based on BERT representation. Thus, even if the correct word is $10$-th hypotheses, our proposal can provide high weight to the correct word.


Attention-based fusion consists of two steps. In the first step, we pick up a representative ASR hypothesis and align the other hypothesis to it. We use the most confident ASR hypothesis, i.e., the 1-best hypothesis, $\hat{S}^1$. If we use multiple ASR systems, we use a hypothesis of a possible best performing ASR system as $\hat{S}^1$. 

We obtain the embedding vectors for the $n$-th hypothesis,  $\tilde{E}^n =[\tilde{e}^n_1, \ldots, \tilde{e}^n_{M^1}] \in \mathbb{R}^{M^1\times B}$, which is time-aligned with the $M^1$-length  hypothesis $\hat{S}^1$, based on the attention mechanism as:
\begin{equation}
\tilde{E}^n=\text{softmax}\left((E^1W^Q) (E^n W^K)^{\mathrm{T}}\right) E^n W^V,
\label{eq:att_time_align}
\end{equation}
$E^n=[e_1^n,\ldots,e_{M^n}^n]^{\mathrm{T}} \in \mathbb{R}^{M^n\times B}$ is the sequence of embedding vectors associated with the $n$-th hypothesis. $E^1$ is the sequence of the 1-best embedding vectors and is used as a query. $\text{softmax}(\cdot)$ is the softmax operation and $W^Q \in \mathbb{R}^{B\times B'}$, $W^K \in \mathbb{R}^{B\times B'}$, $W^V \in \mathbb{R}^{B\times B'}$ are the query, key, and value projection matrices, respectively.

In the second step, we perform attention over the different hypotheses for every aligned sub-word position $m$ in a similar way as the hierarchical attention~\cite{DBLP:conf/naacl/YangYDHSH16,DBLP:conf/interspeech/ManakulGW20,DBLP:conf/icassp/LiuLC19}.
Let $C_{m}=[\tilde{e}^1_{m},\dots , \tilde{e}^N_{m}] \in \mathbb{R}^{B'\times N}$ be a matrix containing the $N$ aligned embedding sequences for the $m$-th sub-word position in a sequence. We can perform attention over the hypotheses to obtain a modified embedding vector $e^{\text{att}}_{m}$ as
\begin{align}
\alpha_{m} = & \text{softmax}\left((e^1_{m})^\mathrm{T} W^Q C_{m} \right), \label{eq:att_weight_hyp}\\
e^{\text{att}}_{m} = &\alpha_{m} C_{m}^\mathrm{T}, \label{eq:att_hyp}
\end{align}
where $\alpha_{m} \in \mathbb{R}^{1\times N}$ are attention weights over the recognition hypotheses.
Eq. (\ref{eq:att_hyp}) performs a similar summation over embedding vectors as in Eq. (\ref{eq:posterior}), but using the attention mechanism to compute the weights and time-aligned embedding vectors.

The proposed attention fusion can also be extended to multi-head attention. In that case, we can obtain an aligned hypothesis for each attention head, $\tilde{E}_h^n$, using a similar equation as Eq. (\ref{eq:att_time_align}), with different projection matrices for each head, i.e., $W^Q_h$, $W^K_h$ and $W^V_h$, where $h$ is the head index.  We can then also define $C_{m,h}$ for each head and compute attention weights $\alpha_{m,h}$ and fused hypotheses ${e}^{\text{att}}_{m,h}$ for each head as in Eqs. (\ref{eq:att_weight_hyp}) and (\ref{eq:att_hyp}). We then obtain the fused hypothesis as,
\begin{align}
    e^{\text{att}}_{m} = & \text{concatenate}(e^{\text{att}}_{m,1}, \ldots, e^{\text{att}}_{m,H})  W^o,
\end{align}
where $H$ is the number of attention heads and $W^o \in \mathbb{R}^{(HB') \times B}$ is an output projection matrix as used in previous work~\cite{DBLP:conf/nips/VaswaniSPUJGKP17}. We used the multi-head implementation in our experiments. Note that we also used the cosine similarity instead of the dot product to compute the attention weights in Eqs. (\ref{eq:att_time_align}) and (\ref{eq:att_weight_hyp}).

\begin{figure}[tb]
 \centering
 \includegraphics[keepaspectratio, scale=.34]
      {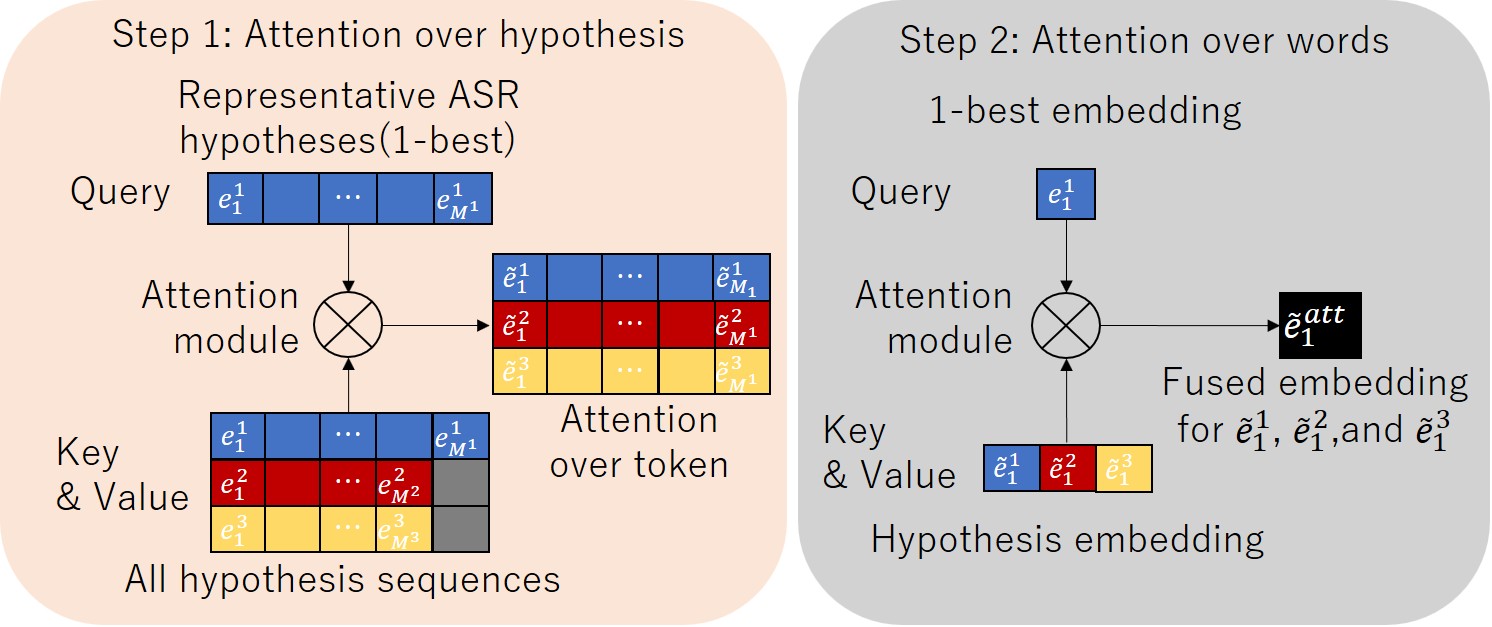}
 \caption{Attention fusion process.}
 \label{fig:MAF}
\end{figure}

Unlike the posterior fusion, we can use hypotheses of different lengths with the proposed attention fusion thanks to the time-alignment step of Eq.(\ref{eq:att_time_align}). Moreover, although we derived the attention fusion assuming it was performed at the input of the BERT encoder, we can also perform attention fusion at any layer within the BERT encoder. When performing the fusion within the BERT encoder, the multiple hypotheses are embedded in the intermediate representation of the BERT model. Therefore, we can exploit BERT's strong language modeling capabilities to combine the hypotheses. 
Figure~\ref{fig:MAF} illustrates how we apply the proposed attention fusion within the BERT encoder.
Note that when we introduce a randomly initialized attention fusion layer, it may corrupt the BERT encoder, making the training slow or unstable. In this paper $B$ equals $B'$, therefore, we initialize the projection matrices $W^Q_h$, $W^K_h$, and $W^V_h$ to identity matrices, so at the beginning of the training, Eq.~\ref{eq:att_hyp} provide highest value to first hypotheses, and the $e^{\text{att}}_{m}$ is close to the 1-best hypothesis $e^1_m$.

\begin{table*}[tb]
\centering
 \small
 \caption{Comparison of the text and speech summarization corpora.}
 \label{tb:dataset}
  \begin{tabular}{lccccccccc}
  \toprule
    Dataset & Text  & Speech  & \multicolumn{2}{c}{Compression rate}& \multicolumn{2}{c}{Source lengths}  & \multicolumn{2}{c}{Target lengths}  & WER\\
  & documents & documents & sentence & word & sentence & word & sentence & word &  \\ \midrule
    CNNDM & 162,018 & n/a & 14\% & 9\%  & 35& 853   & 3& 60 & n/a \\ 
    How2 & 72,983 & 12,798  & 16\% & 16\% & 14& 303   & 1& 34 & 13.0\% \\ 
    TED & 4,001 & 1,495 & 6\% & 5\%   & 102& 2210 & 4& 79 & 8.5\% \\\bottomrule
  \end{tabular}
\end{table*}

\vspace{-7pt}
\section{TED Speech summarization corpus}
\vspace{-7pt}
\label{sec:data}
We used three corpora to train and test our speech summarization systems. In particular, we assembled a new corpus derived from TED Talks that we will release upon acceptance of the paper. Table~\ref{tb:dataset} compares the characteristics of the TED corpus with the two other corpora used in this paper.
\vspace{-6pt}
\subsection{Descriptions of the corpora}
\vspace{-5pt}
CNN-DailyMail (CNNDM) is a large-volume corpus used for text summarization of news documents. We use this corpus to pre-train the TS system. How2 is a publicly available corpus for speech summarization~\cite{DBLP:journals/corr/abs-1811-00347}. It consists of summarizations of How2 videos taken from YouTube. The target for summarization consists of the brief video description provided on YouTube. Although the corpus also includes videos, allowing multi-modal summarization, here we only use the audio content of the corpus. 

The TED corpus consists of a summarization of TED Talks.  We created this corpus by associating TED Talks included in the TEDLIUM corpus with their summaries obtained from the TED website\footnote{https://www.ted.com/talks}. For the TED summarization task, we add the speaker name to the speech document to allow the TS systems to output the speaker names, which commonly appear in the reference summaries. The target summary consists of the title and abstract of the talk. Note that others have also used TED Talks from speech summarization~\cite{DBLP:conf/apsipa/KotoSNTAN14}, but those corpora were small. Furthermore, they did not publicly release their corpus.
\vspace{-10pt}
\subsection{Analysis of the complexity of the TED summarization task}
\vspace{-5pt}
Table~\ref{tb:dataset} shows the number of text and speech documents, the compression rate, the source and target lengths, and the word error rate (WER) of the three corpora. 
The compression rate is expressed as the ratio between the output and input document lengths. Thus, a lower value means higher information compression.
This table shows that our proposed TED summarization task is challenging because it has a relatively small amount of training data and consists of long input speech documents that require higher information compression than needed for existing datasets. Note that the WER is about 8.5\%, which makes it slightly easier than the How2 corpus in terms of the ASR performance.

\begin{table}[tb]
\centering
 \small
 \caption{ROUGE scores of ideal extractive summaries and word overlap.}
 \label{tb:extractive}
  \begin{tabular}{lcccc} 
  \toprule
    Dataset & ROUGE-1 & ROUGE-2 & ROUGE-L & Word overlap  \\ \midrule
    CNNDM   & 45.0 & 29.9 & 42.8 & 83\%   \\ 
    How2    & 27.9 & 10.4 & 23.3 & 58\%     \\ 
    TED     & 34.4 & 19.8 & 33.5  & 72\%  \\ \bottomrule 
  \end{tabular}
\end{table}

Another way in analyzing the complexity of a summarization task is whether it is to apply extractive summarization. We can measure ideal extractive summarization scores using the reference summary to select the set of sentences from the input document that achieves the highest summarization score. Table~\ref{tb:extractive} shows the ROUGE scores~\cite{lin-2004-rouge} and word overlap of the ideal extractive summary for the three datasets. The word overlap measures the percentage of words from the target summaries that are in the source documents.
CNNDM, which consists of a summary of news articles, is well suited for extractive summarization; therefore, oracle scores and word overlap are relatively high. In contrast, the How2 corpus consists of relatively casual speech, which leads to much lower oracle scores and word overlap. The TED summarization corpus consists of relatively formal speech and its oracle extractive summarization scores are between those of CNNDM and How2.

\begin{table}[tb]
\centering
 \small
 \caption{ROUGE scores of  abstractive summarization with BERTSum using the ground-truth transcriptions (BERTSum (oracle)).}
 \label{tb:abstractive_top}
  \begin{tabular}{lccc} 
  \toprule
    Dataset & ROUGE-1 & ROUGE-2 & ROUGE-L   \\ \midrule
    CNNDM   & 41.7 & 19.4 & 38.8    \\ 
    How2    & 56.5 &37.8 &59.3    \\ 
    TED     & 32.1 &6.2 & 19.0    \\ \bottomrule 
  \end{tabular}
\end{table}

Finally, we compare the performance of abstractive text summarization on the ground-truth transcriptions. The results indicate an upper-bound value for the speech summarization systems we investigated. Table~\ref{tb:abstractive_top} shows the ROUGE scores obtained with BERTSum models for the three tasks. 
We found that the difficulty of abstractive summarization is highly dependent on the length of the input and the compression rate, which makes the proposed TED summary task also challenging for abstractive summarization.

Comparing the results of Table~\ref{tb:extractive} and \ref{tb:abstractive_top} shows that ideal extractive summarization's scores on TED talk, in particular ROUGE-2 and -L scores, are higher than abstractive summarization ones. We will investigate extractive summarization on the TED corpus and comparing it with abstractive summarization in future works.

\begin{table}[tb]
\centering
 \small
 \caption{ASR WER for each BPE size. 30.5k$^*$ corresponds to the BPE size of the BERT model. }
 \label{tb:wer}
  \begin{tabular}{lcccccc} 
  \toprule
  & \multicolumn{6}{c}{BPE size} \\
     &  500 &  5k & 10k &  20k  &  30k &  30.5k$^*$ \\ \midrule
    How2    & n/a     & 13.0   & 13.6    & 14.1    & 14.3     & 14.6       \\ 
    TED     & 8.5     & n/a    & 8.7    & 9.5      & 10.0    & 10.4     \\\bottomrule
  \end{tabular}
\end{table}

\vspace{-10pt}
\section{Experiments}
\vspace{-8pt}
\label{sec:Experimental setup}
We performed experiments using two speech summarization datasets, TED and How2 described in Section~\ref{sec:data}.
\vspace{-10pt}
\subsection{System configuration}
\vspace{-5pt}
Our baseline consists of a cascade of ASR and TS systems trained separately.
We built Transformer-based ASR models using the Espnet toolkit\footnote{https://github.com/espnet/espnet}, by following the published recipes for the TEDLIUM2~\cite{DBLP:conf/lrec/RousseauDE12} and How2~\cite{DBLP:journals/corr/abs-1811-00347} tasks, except that we varied the BPE size.
For the TS model, we built a BERTSum model using a pre-trained BERT model as an encoder and a Transformer decoder in the same way as~\cite{DBLP:conf/icassp/LiuLC19}. 
We used the pre-trained BERT model provided by huggingface\footnote{https://huggingface.co/transformers/}. We pre-trained the BERTSum model using CNNDM data, and fine-tuned it on each summarization task. 

We consider three baseline systems, ``(1) BERTSum (oracle),'' which uses ground-truth ASR transcriptions as input to the BERTsum model, ``(2) BERTSum (ASR-BPE),'' which uses transcriptions obtained with an ASR system trained with the optimal BPE definition for ASR, and ``(3) BERTSum (BERT-BPE),'' which uses an ASR system trained with the BPE definition of the pre-trained BERT model.
The first baseline illustrates the upper-bound performance on the tasks. The second baseline is optimal for ASR, but it requires re-tokenizing the recognized text to match the BPE of the BERTSum model, and it cannot be used to pass ASR information such as confidence or posteriors to the summarization back-end. The third baseline uses the same BPE definition as our proposed methods.

\begin{table*}[htb]
\centering
 \small
 \caption{ROUGE scores for the different speech summarization systems. Systems (6) and (7) are the proposed methods. }
 \label{tb:scores}
  \begin{tabular}{llcccccc} 
  \toprule
   & Method     & \multicolumn{3}{c}{TED}                    & \multicolumn{3}{c}{How2}                     \\ 

    && ROUGE-1 & ROUGE-2 & ROUGE-L & ROUGE-1 & ROUGE-2 & ROUGE-L \\   \midrule    
    (1) &BERTSum (oracle)  & 32.1 &6.2 & 19.0          &  56.5 &37.8 &59.3 \\ \midrule
    (2) &BERTSum (ASR-BPE) & 29.9 &6.9 &18.3 &  47.4 &27.1 &46.1 \\ 
    (3) &BERTSum (BERT-BPE) & 28.9 &6.2 &17.8  &     45.3&26.8 &45.0  \\ 
    
    (4) &BERTSum retrain & 31.5 &5.6 &20.4     &   47.2 &27.0 &45.6 \\  
    (5) &BERTSum confidence & 30.1 &6.8 &20.4   &     48.4 &29.0 &47.3   \\ \midrule 
    (6) &BERTSum Pos. fusion  & 31.6 &6.1 &20.3    & 48.2 &27.8 &46.5  \\ 
    (7) &BERTSum Att. fusion & 31.9 &6.0 &19.3 &  49.3 &28.8 &48.2 \\ 
    \bottomrule
  \end{tabular}
\end{table*}

In addition to the above systems, we created a baseline by retraining (3) BERTSum (BERT-BPE) on the transcriptions generated with ASR (``(4) BERTSum retrain''). We also implemented a system similar to~\cite{DBLP:conf/eusipco/WengLC20} that uses BPE confidence scores as an auxiliary feature at the input of BERTSum (``(5) BERTSum confidence'') as described in Section~\ref{sec:confidence}. 

\begin{table*}[tb]
\centering
 \small
 \caption{How2 dataset summarization examples. The red collar highlights the different main parts. }
 \label{tb:summary}
  \begin{tabular}{cp{13.5cm}} 
  \toprule
  Method & \multicolumn{1}{c}{Example} \\ \hline

Reference& \textcolor{red}{learn how to form a b sound for ventriloquists} with expert voice throwing tips from a professional comedian in this free online ventriloquism lesson video clip \\ \hline

(5) BERTSum confidence & \textcolor{red}{practice your ventriloquists} with expert voice throwing tips from a professional comedian in this free online ventriloquism lesson video clip \\ \hline

(7) BERTSum Att. fusion &\textcolor{red}{learn how to make b sound for ventriloquists} with expert voice throwing tips from a professional comedian in this free online ventriloquism lesson video clip \\ \toprule

Reference & when choosing the right hair style for your face , pull your hair back and \textcolor{red}{take into account the shape of your face} choose the right hair style with tips from a beauty professional in this free video on hair care \\ \hline

(5) BERTSum confidence & when picking a hairstyle for face , \textcolor{red}{it ' s important to pick a hairstyle that is n ' t fit} choose a hairstyle with tips from a beauty professional in this free video on hair care \\ \hline

(7) BERTSum Att. fusion & when choosing a flattering hairstyle , \textcolor{red}{take an measurements of the face shape} and pull all of the hair back into account choose a flat hairstyle with tips from a beauty professional in this free video on hair care \\ \bottomrule
  \end{tabular}
\end{table*}

The proposed posterior-based hypotheses fusion used the same ASR and TS system as the baseline systems, i.e. (3) BERTSum (BERT-BPE), except that the BERTSum system was fine-tuned on input embeddings obtained with Eq. (\ref{eq:posterior}) using $N{=}10$ as described in Section~\ref{sec:posterior fusion}. We call this system ``(6) BERTSum Pos. fusion''.

For the proposed attention-based fusion, we inserted the attention fusion layer at the fifth layer of the BERT and used four attention heads ($H=4$).
We used the same approach to generate the recognition hypotheses as for the posterior fusion described in Section~\ref{sec:posterior fusion} except that we used attention fusion described in Section~\ref{sec:attention fusion} instead of Eq. (\ref{eq:posterior}). We used five hypotheses ($N{=}5$) for attention fusion due to GPU memory constraints of our experimental environment. We trained all BERTSum models following the original recipe except that we used a learning rate of $0.0002$ and warm-up steps of $20$k when retraining on the ASR outputs (systems (4) to (7)). We compared the summarization performance of each method with ROUGE~\cite{lin-2004-rouge}. 

\vspace{-8pt}
\subsection{Effect of BPE size}
\vspace{-5pt}
\label{sec:bpe size}
ASR and BERT models use both BPE to represent sub-words; however, the BPE unit definitions used for the two systems differ. Typically, ASR systems achieve optimal performance at a smaller BPE size than that of the BERT model. 
However, since our proposed method requires that the BPE unit definition of the ASR system must match that of the BERT model, we expect to have some ASR degradation due to too large BPE sizes. Thus, we first investigate the effect of BPE size on ASR performance.

Table~\ref{tb:wer} shows WER as a function of different BPE sizes for the How2 and TED corpora. We observe a relative WER increase of 12\% for How2 and by more than 20\% for the TED corpus when adopting the BPE definition used by BERT. Although this is a significant WER increase, we discuss its impact on summarization in the following subsection.

\vspace{-8pt}
\subsection{Speech summarization results}
\vspace{-5pt}
\label{sec:Experiments}

Table~\ref{tb:scores} shows the ROUGE scores for the baseline systems (systems (1) to (5)) and the proposed method with (6) posterior and (7) attention-based fusion\footnote{Note that we confirmed the validity of our implementation of BERTSum as it achieved a similar level of performance on the How2 corpus~\cite{DBLP:conf/emnlp/LiuL19}, which reported  ROUGE-1 and ROUGE-L scores of 48.3 and 44.0, respectively, although the systems cannot be directly compared because of differences in the training data and ASR front-end.}.
\begin{figure}[tb]
 \centering
 \includegraphics[keepaspectratio, scale=.27]
      {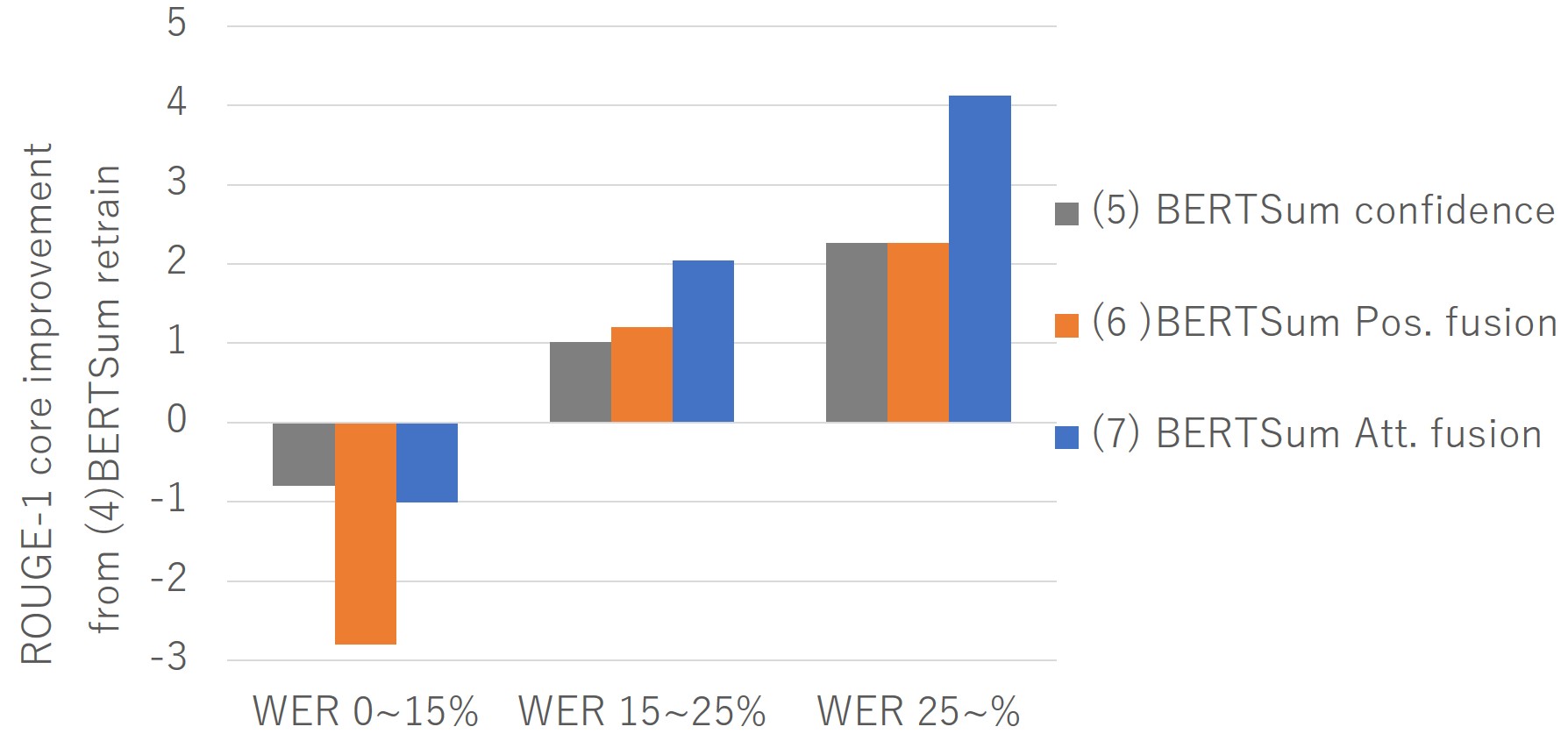}
 \caption{ROUGE score improvement from (4)BERTSum retrain for (5) BERTSum confidence and the proposed methods as a function of the WER of the input documents.}
 \label{fig:diff}
\end{figure}
We observe a large performance gap in summarization performance when using ASR transcriptions (system (2)) instead of ground-truth transcription (system (1)).  
As we discussed in section~\ref{sec:bpe size} using BERT's BPE definition for ASR (system (3)) induces more recognition errors, which clearly degrades summarization performance on both tasks compared to using the BPE optimal for ASR (system(2)). However, this performance degradation can be mostly recovered by fine-tuning on the ASR hypotheses (system (4)).
Using confidence scores (system (5))~\cite{DBLP:conf/emnlp/LiuL19} improves ROUGE scores on the How2 corpus, but degrades ROUGE-1 on the TED corpus compared to simply retraining on the ASR hypotheses.

The proposed posterior (system (6)) and attention-based fusion (system (7)) approaches both achieve equivalent or superior ROUGE-1 scores for both corpora than the baseline systems with retraining on ASR hypotheses (system (4)). In particular, the proposed attention-based fusion improves ROUGE-1 by 2 points on the How2 corpus. This result indicates that the proposed system can better mitigate ASR errors by using multiple hypotheses with the BERTSum model.

Figure~\ref{fig:diff} shows the ROUGE-1 score difference between systems (4) and (5)-(7) as a function of the WER of the spoken documents. These results show that our proposed attention fusion system achieves higher robustness for ASR error than systems (5) and (6). 
We hypothesize that this is due to the fact that the proposed attention fusion approach models explicitly multiple ASR hypotheses, and can thus choose alternative word candidates within ASR hypotheses to generate the summary. This may make the attention fusion-based system robust to ASR errors if the correct words are included in the hypotheses. In addition to the ROUGE score, Table~\ref{tb:summary} provides a couple of summaries generated by our proposed system for the How2 corpus~\footnote{We provide more examples, as well as examples on the TED corpus on our webpage \url{https://github.com/takatomokano/ted_summary}}. We include summaries obtained with BERTSum fine-tuned on the ASR hypotheses as a comparison. We confirm that both systems achieve highly readable summaries that are close to the reference.

The experiments with the proposed attention fusion used aligned hypotheses of the same duration. However, the proposed method can also handle hypotheses of different lengths thanks to the alignment mechanism of Eq. (\ref{eq:att_time_align}). We also tested the attention fusion using hypotheses generated by five different ASR systems each with a different BPE size (the systems used in the experiments of Section~\ref{sec:bpe size}). In this case, the hypotheses were unaligned and of different lengths. The proposed attention fusion achieved a ROUGE-1 score of 48.0 on the HOW2 task, which shows some improvement over the baseline systems (2)-(4). Although this result is behind our best performing system, it shows that the proposed method could handle hypotheses with different lengths. We plan to further investigate such a combination by using more diverse ASR systems to generate the hypotheses in future work.

\vspace{-19pt}
\section{Conclusion}
\vspace{-8pt}
In this paper, we proposed a speech summarization system that can exploit multi-hypotheses generated by an ASR system to mitigate the impact of recognition errors. We proposed two schemes, i.e., posterior and attention-based fusion, which could be integrated into a BERT-based TS model. We showed that both approaches could reduce the impact of ASR errors on summarization and achieved competitive results on two tasks.

Future works will include investigating tighter interconnection of the ASR front-end and TS back-end to mitigate ASR errors and exploit speech-specific information such as intonation and create richer and more informative speech summaries.

\vspace{-10pt}
\section{Acknowledgement}
\vspace{-8pt}
We would like to thank Jiatong Shi at Johns Hopkins University for providing a script of ROVER-based system combinations.

\bibliographystyle{IEEEbib}
\bibliography{strings,refs}

\end{document}